# A Digital Spreading Framework for Quantum Expectation Computation Without Rotation Gates or Arithmetic Circuits

Yu-Ting Kao
Industrial Technology Research Institute
Hsinchu, Taiwan
SunnyKao@itri.org.tw

Yeong-Jar Chang
Industrial Technology Research Institute
Hsinchu, Taiwan
ot@itri.org.tw

**Abstract**—In the pursuit of quantum advantage for financial engineering, researchers face a critical dilemma: analog rotation gates suffer from inherent 'sine-to-square' biases and error magnification, while digital arithmetic circuits (e.g., WeightedAdder) incur prohibitive quadratic complexity that exceeds NISQ capabilities. This study introduces Digital Spreading (DS), a fully digital quantum computing framework designed to resolve this trade-off. DS overcomes these limitations by utilizing a pruned Cuccaro ripple-carry architecture that avoids costly multiplication and eliminates rotation gates entirely. The proposed circuit employs integer comparison operations on superposed quantum states, mapping multi-qubit outcomes onto the probability of a single target qubit. Experiments based on a random walk model for option pricing demonstrate that DS achieves floating-point precision with a relative error as low as 0.0001%, outperforming JP Morgan's rotation-based method (1.43%) [1], as well as ITRI's analog calibration (1.43%) and digital calibration approaches (19.14%) [2]. Overall, DS provides a compact, robust, and accurate framework for quantum weighted-average computation.

## I. Introduction

Although quantum computing offers the potential for massive parallelism, achieving quantum advantage remains challenging without suitable algorithms. Brassard et al. [4] proposed Quantum Amplitude Estimation (QAE), providing a theoretical framework for estimating expected values in exponential-scale computations with a convergence rate of $O(1/m)$, thereby establishing a quantum advantage over classical Monte Carlo methods. Building on this, Montanaro [5] analyzed the systematic quantum advantage of Monte Carlo approaches and highlighted potential application scenarios where QAE could yield significant speedups, further demonstrating the practical relevance of quantum acceleration in Monte Carlo–based computations.

Traditional Monte Carlo simulations are fundamental for financial modeling but require extensive random sampling and statistical averaging, which makes efficient implementation on quantum hardware challenging. Noto [6] introduced novel analog circuit designs specifically for estimating the value of $\pi$ using QAE, while ITRI [7] presented pioneering digital Monte Carlo circuit designs that faced exponential complexity growth due to reliance on truth tables and lacked quantum advantages without integrating QAE. These Monte Carlo–based computations extend quantum applications beyond finance to a wide range of industries, including option pricing and risk analysis in finance, molecular simulation in chemistry, uncertainty quantification in engineering, and stochastic modeling in logistics and energy systems.

In many of these applications, a central operation is the computation of weighted averages over probability distributions encoded in quantum states, a key step in Monte Carlo simulations, expected value estimation, and derivative pricing. Probability mass functions (PMFs) can be derived from mathematical models or obtained from statistical results of direct Monte Carlo simulations. The weighted-average computation circuit offers two inherent advantages. First, floating-point probabilities can be encoded across the amplitudes of a $2^n$-qubit state, allowing quantum measurement to operate on these probabilities far more efficiently than conventional arithmetic circuits [8]. Second, multi-bit outputs can be merged into a single qubit, enabling the use of a single QAE instance rather than multiple QAEs.

Existing approaches for weighted-average computation in quantum finance include rotation-gate–based methods and Qiskit WeightedAdder (QWA). Rotation-based methods encode probabilities into qubit amplitudes via analog rotations, but they are sensitive to noise and introduce sine-to-square approximation errors. On the other hand, QWA implements arithmetic digitally but suffers from significant circuit overhead. These limitations restrict the scalability of current techniques and hinder their deployment on near-term quantum hardware, especially when high precision or large problem sizes are required.

In this work, we propose DS, a fully digital quantum framework that addresses these challenges by leveraging value-to-state spreading. For example, a value of 5 can be encoded as 11111000 across 8 quantum states, representing 5 ones and 3 zeros. This approach eliminates the need for rotation gates. By mapping multi-qubit outcomes onto a single target qubit through comparison operations, DS reduces circuit complexity and improves numerical stability, providing a compact and robust method for weighted-average computation. The remainder of this paper is organized as follows: Section 2 reviews related work in quantum finance and Monte Carlo–based algorithms, Section 3 presents the DS framework, Section 4 evaluates its performance in option pricing simulations, and Section 5 concludes with discussions on future directions.

## II. Related Works

In the domain of Monte Carlo-based applications, the quantum circuit can be generally structured into three distinct phases(**Fig.1**): In general, the Broadcast module is responsible for generating superposition quantum states; the Calculation module performs parallel evaluation of target functions; and the Measurement module extracts useful outcomes via efficient algorithms. In the present case, Broadcast prepares the probability distributions, Calculation computes the product of the stock prices and their corresponding probabilities parallelly, and Measurement employs QAE to obtain the overall expected value. The proposed DS is designed to replace the existing Calculation module, which is responsible for computing the weighted average of a random variable defined over a discretized probability distribution in the context of quantum financial modeling, such as option pricing.

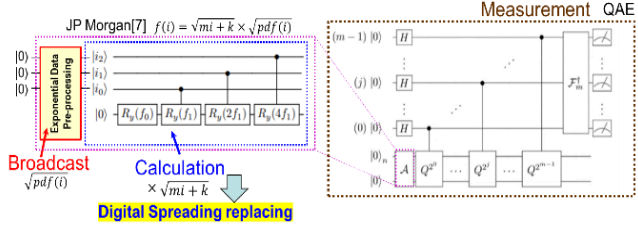

**Fig. 1**. Digital Spreading replaces the Calculation module

Consider a 3-qubit system representing 8 possible states ($N = 8$), with a probability distribution $P$ and associated state values $V$ (e.g., asset prices or indices). A illustrative scenario is defined as follows:

$$P = [0, 0.25, 0.25, 0, 0.25, 0.25, 0, 0]$$

Here, non-zero probabilities exist only at indices $|1\rangle, |2\rangle, |4\rangle$, and $|5\rangle$. Target Weighted Average (WAG) can be calculated as:

$$WAG = p_i \cdot i = 0.25(1) + 0.25(2) + 0.25(4) + 0.25(5) = 3$$

To address the problem, the existing Calculation module can be broadly categorized into two approaches: Analog Rotation–based Approaches and Digital Arithmetic Approaches, which are described as follows.

*A. Analog Rotation-based Approaches*

The prevailing methods for WAG calculation, notably those pioneered by J.P. Morgan [1] as shown in **Fig 2**. After the rotation operations, the amplitude of each state is given by

$$\sqrt{P_i} \cdot \sin(\frac{\pi}{4} + i\theta)$$

where $i$ represents the quantum states 000, 001, 010, ... 111, $P_i$ is the corresponding probability, and $\theta$ is a small fundamental rotation angle.

All information is consolidated into a single quantum "Analog" register. Quantum Amplitude Estimation (QAE) is employed to extract the probability of this register. The probability of measuring a particular basis state is given by the square of the magnitude of its amplitude. Thus, the probability of observing a specific measurement outcome for the "Analog" qubit is given by the sum of the probabilities of all constituent basis states. In the following formula, QAE denotes the probability estimated via the QAE procedure, rather than the algorithm itself:

$$QAE = \sum P_i \cdot \sin^2(\frac{\pi}{4} + i\theta)$$

By applying a Taylor series approximation, the square can be linearized:

$$\sin^2(\frac{\pi}{4} + i\theta) \approx \frac{1}{2} + x$$

yielding:

$$QAE \approx \sum P_i \cdot (\frac{1}{2} + i\theta) = \frac{1}{2} + \sum P_i \cdot (i\theta)$$

For $\theta = 0.01$, the weighted average of interest can be expressed as:

$$WAG = \sum P_i \cdot i \approx 100 \cdot (QAE - 0.5)$$

If $QAE = 0.53$, the weighted average (WAG) would ideally be 3, as described above. However, the Taylor series approximation used in the "sin-to-square" mapping introduces a non-negligible error. In the Qiskit simulation, QAE yields 0.52957, resulting in a WAG of 2.957. At first glance, the Taylor series approximation appears to introduce a negligible error, as evidenced by the comparison between 0.53 and 0.52957. However, during the conversion from the quantum measurement result (QAE) to the weighted average (WAG), the subtraction of 0.5 in the calculation reduces the effective denominator, thereby amplifying the relative error. As a result, the final comparison between 3 and 2.957 exhibits a larger relative discrepancy (1.43%).

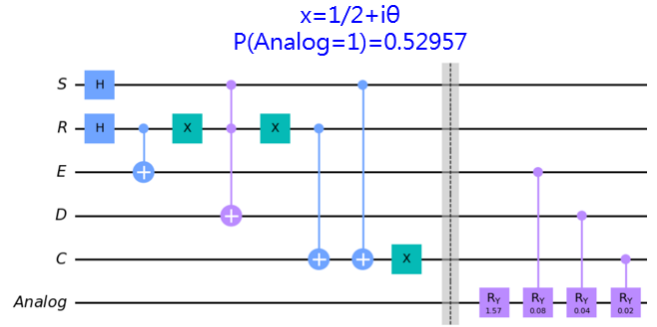

**Fig. 2**. JP Morgan's Rotation-based circuit suffers from 1.43% error due to Taylor series approximation.

Following the work of JP Morgan, ITRI [2,3] proposed two weighted-average computational circuits, namely Analog Calibration and Digital Calibration, both of which build upon and improve the original method. Using the same example as above, the quantum circuit designer aims to vary the rotation angle y around π/4; accordingly, the initial stock price x is also set to π/4:

$$x = \frac{\pi}{4} + i\theta$$

$$y = \frac{\pi}{4} + i\theta$$

As shown in **Fig 3** (left), QAE yields an estimated probability of 0.54697 due to the increased computation error, resulting in WAG = 4.597. The error 3.202% is larger than the value claimed by JP Morgan, which is attributed to the initial stock price being $\pi/4$ rather than $1/2$. Analog Calibration is a correction method applied when the initial point deviates from $1/2$, in which the stock price increment is scaled by a factor of m, as follows:

$$x = \frac{\pi}{4} + mi\theta$$

$$y = \frac{\pi}{4} + i\theta$$

When $m = \pi/2$, the error is minimized, as shown in **Fig 3** (right), yielding an error 1.43%, identical to the original JP Morgan result. Mathematically, these two expressions differ only by a linear transformation, which explains why the errors are equal:

$$x = \frac{\pi}{4} + \left(\frac{\pi}{2} \cdot i\theta\right) = \frac{\pi}{2} \cdot \left(\frac{1}{2} + i\theta\right) = \frac{\pi}{2} \cdot \sum_{n=0}^{\infty} \frac{f^{(n)}(a)}{n!} (i\theta)^n$$

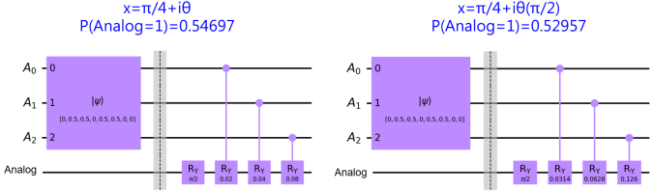

**Fig. 3**. ITRI's analog calibration circuit reduces the error from 3.202% to 1.43% via different stock price increment scenarios

For a general system $y = f(x)$, where x is a digital input and $y$ is an analog output, output errors can be compensated by adjusting the digital input to control the discrete levels such that the resulting output matches the expected value. This approach is referred to as Digital Calibration, which constitutes the second method proposed by ITRI [2].

In the financial application considered in this work, errors primarily arise from the sin-to-sqrt transformation. To mitigate this issue, an additional design block is introduced to perform a digital transformation defined as

$$r = arcsin(\sqrt{x})$$

During the quantum rotation operation, the output is given by

$$\sin(r) = \sin(arcsin(\sqrt{x})) = \sqrt{x}$$

thereby achieving an exact sin-to-sqrt conversion.

However, a limitation of this approach is that $r$ is represented as a digital signal. When the number of qubits is insufficient, quantization error becomes significant. Consequently, in the case shown in **Fig. 4**, the overall error reaches 19.14%.

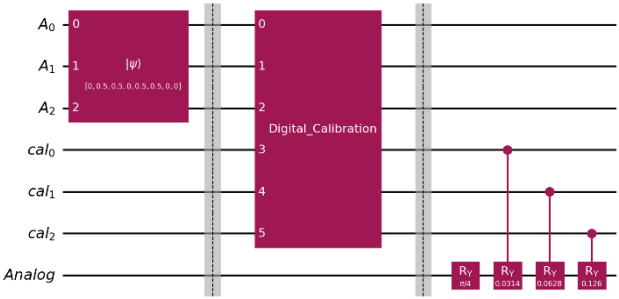

**Fig. 4**. ITRI's Digital Calibration circuit exhibits a 19.14% quantization error due to the limited bit-width in the calibration module.

### B. Digital Arithmetic Approaches

To circumvent analog approximation biases, purely digital alternatives such as Qiskit WeightedAdder (QWA) have been employed. This approach enables integer cost function calculations across quantum superpositions without introducing sine-to-square bias, theoretically providing higher precision (**Fig. 5**). In addition, QWA allows non-binary weighting for individual qubits, a capability absent in previous implementations by JP Morgan and ITRI, which were limited to powers-of-two weight increments (1, 2, 4, 8, …), thereby restricting stock price variations to linear increments. By supporting arbitrary integer weight changes, QWA extends its applicability to a wider range of problems, including portfolio optimization, combinatorial optimization such as knapsack problems, and multi-objective function evaluation.

Nevertheless, QWA exhibits quadratic circuit complexity and considerable ancilla qubit requirements, which constrain its scalability on NISQ (Noisy Intermediate-Scale Quantum) devices. While previous research has extensively explored constant-optimized quantum circuits to reduce the gate count and depth of modular multiplication [9], explicit arithmetic operations inherently demand substantial hardware resources. Furthermore, the circuit does not consolidate multi-qubit outputs into a single qubit, necessitating multiple QAE instances to realize quantum advantage (**Fig. 6**). These factors pose significant challenges for practical deployment.

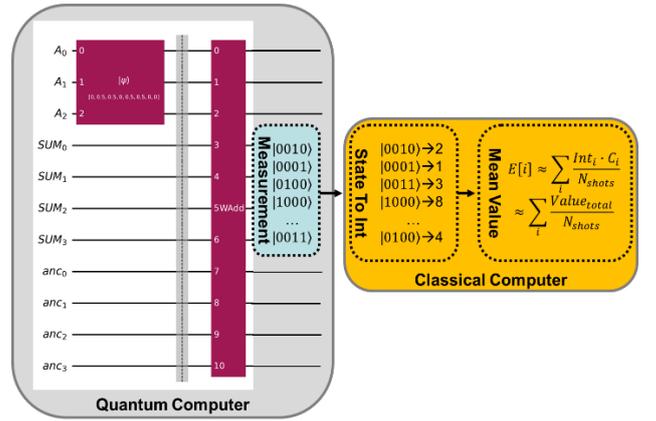

**Fig. 5.** QWA circuit and mean-value computation flow

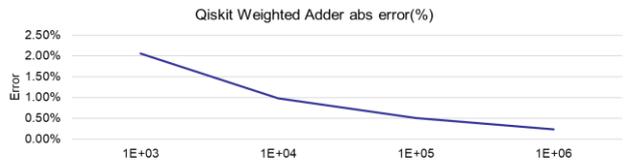

**Fig. 6**. QWA requires 10,000,000 shots to improve the error rate to 0.0315%.

Finally, Table I summarizes the comparison between related works and the proposed Digital Spreading (DS) approach. Regarding the gate requirements, JP Morgan [1] and ITRI [2] both employ Ry gates, whereas QWA and DS do not, allowing the methods to be classified as analog and digital, respectively. All approaches support the important feature of converging to a single qubit except QWA. In terms of robustness, digital methods outperform analog ones. Concerning the rotation angle and initial price requirements in [1], analog calibration relaxes the initial price constraint, while digital calibration alleviates both constraints.

*Table I* Related Work

| Weighted Average Computation | Rotation-Based Methods | | | Digital Methods | |
|---|---|---|---|---|---|
| | JP[1] Rotation gate | ITRI[2] Analog Calibration | ITRI[2] Digital Calibration | QWA | Proposed DS |
| Ctrl-Ry gates required | O | O | O | X | X |
| Hadamard gates required | X | X | X | X | O |
| Digital gates required (X, CX, CCX) | X | X | O | O | O |
| Converge to single qubit | O | O | O | X | O |
| Robustness | Larger error due to Taylor approximation on rotation-gate | | | Good | Good |
| Rotation angle required | 45º | 45º | Calibrated by digital LUT | No limitation | No limitation |
| Asset's initial price required | 1/2 | Calibrated by m/2+mx | | No limitation | No limitation |

## III. METHODOLOGY

This section presents the algorithmic architecture and circuit implementation of the Digital Spreading (DS) framework, a fully digital computational paradigm for weighted-average computation designed to mitigate the inherent inaccuracies introduced by rotation-based methods due to Taylor-series approximations, as well as the prohibitive resource overhead associated with arithmetic operations implemented using the Qiskit WeightedAdder (QWA).

To enable measurement and reduce the required Quantum Amplitude Estimation (QAE) instances from multiple to a single one, the circuit needs to transform the result of a multi-bit quantum computation into the probability of a target qubit. The proposed DS implements this transformation by leveraging quantum superposition and an overflow-based evaluation mechanism, rather than relying on the rotation-gate approach proposed by JP Morgan.

Furthermore, the internal circuit structure of the DS framework employs pruned ripple-carry operations to preserve numerical precision while significantly reducing circuit depth and gate complexity compared with conventional quantum floating-point arithmetic approaches. In this design, floating-point values are encoded directly in the amplitudes of Dirac states, allowing the weighted average of stock prices under different probabilities to be mapped onto the probability of a target qubit without introducing numerical errors. In contrast, computing the weighted average through explicit quantum arithmetic would require a large number of gates and may introduce floating-point rounding errors.

### A. Digital Probability Encoding via Overflow Logic

The proposed DS utilizes a maximally superposed ancillary register—functioning as a "Digital Ramp"—to directly map multi-bit integer sums onto the probability space of a single target qubit (Fig. 7).

Within the DS framework, the expected value of a discrete variable $A$ (bounded by $N=2^n$) is considered. Since $A$ is defined over a finite set of discrete outcomes, the probability density function (PDF) is replaced by the corresponding probability mass function (PMF). Accordingly, the expected value is computed as a probability-weighted sum:

$$WAG = \sum_x P(x) \cdot A_x = N \sum_x P(x) [\frac{1}{N} \sum_{K=0}^{N-1} (A_x > K)]$$

Aanalytically, evaluating the magnitude of $A$ is equivalent to determining the probability that $A$ exceeds a uniformly distributed random threshold $K \in [0, N-1]$. To efficiently realize this within NISQ constraints, the DS architecture evaluates the condition $A + K' \geq N$ (where $K' = N - 1 - K$) by adding the asset value $A$ to the Digital Ramp $K'$ and observing only the overflow qubit, which is equivalent to $A > K$. This operation effectively spreads the cumulative distribution of the multi-bit outcome into the amplitude of a single ancilla qubit:

$$|A\rangle \otimes \sum_{K=0}^{N-1} |K\rangle \otimes |0\rangle \xrightarrow{DS} |A\rangle \otimes \sum_{K=0}^{N-1} |K\rangle \otimes |A > K\rangle$$

Because the threshold is uniformly distributed, the measurement probability of this ancillary qubit in the active state is strictly linearly proportional to the macroscopic expectation value of $A$.

From an algorithmic fidelity perspective, this deterministic linear mapping ($P \propto A$) completely bypasses the intrinsic non-linear degradation ($P \propto \sin^2(\theta)$) of conventional Taylor series approximation [1] and calibration-based corrections, including analog spacing adjustment and digital arcsine transformation [2].

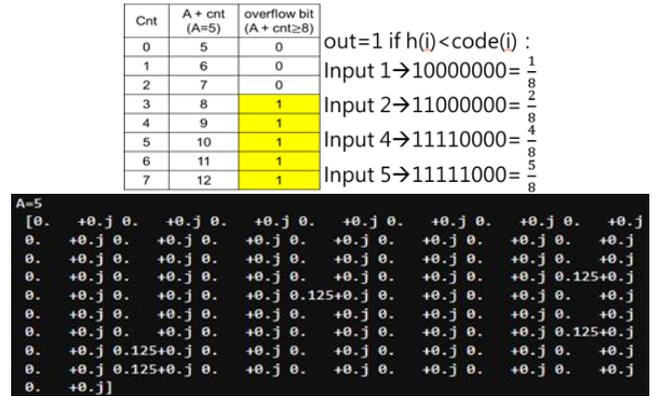

**Fig. 7**. Value-to-state spreading: Since the addition A + cnt results in five overflow cases, the state vector contains five nonzero terms.

By maintaining global unitarity and minimizing intermediate entanglement entropy, the DS mechanism confines system inaccuracy to statistical shot noise. This enables exact encoding of weighted averages without analytical approximation bias or explicit arithmetic operations. Furthermore, it supports parallel evaluation of discrete cost functions over the full $N = 2^n$-dimensional Hilbert space while preserving relative phase information for subsequent amplitude amplification.

### B. Circuit Architecture with the Pruned Cuccaro Adder

The physical implementation of this Overflow Logic relies on Cuccaro ripple-carry Majority (MAJ) elements[10] as shown in **Fig 8**. Unlike the Qiskit WeightedAdder (QWA), which suffers from quadratic gate scaling ($O(n^2)$) and requires extensive auxiliary registers, the proposed Cuccaro-centric topology strictly confines circuit depth to a linear $O(n)$ growth model. This linear dependency mitigates noise accumulation and gate infidelities, rendering the architecture exceptionally scalable for decoherence-limited NISQ processors.

To implement $A + K' \geq N$, the stock price value $A$ is generated through arithmetic operations or Monte Carlo simulations on the input registers $A_0$, $A_1$, and $A_2$. The digital ramp $K'$ is implemented using auxiliary counters $cnt_0$, $cnt_1$, and $cnt_2$, and the final summation $A + K'$ is realized using the pruned Cuccaro adder. During this process, intermediate carry signals propagate coherently along the qubit chain, while transient ancillary registers are retained only for the minimal duration required to support the ripple-carry propagation. Since the objective is solely to detect the overflow condition, the circuit can be significantly simplified. Specifically, only the intermediate ripple-carry propagation needs to be evaluated, while the explicit computation of the partial sums is unnecessary. This optimization effectively eliminates approximately half of the adder circuitry, reducing the overall gate complexity.

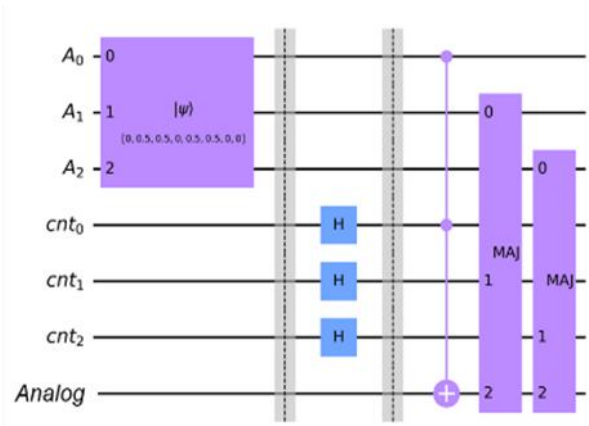

**Fig. 8**. Digital Spreading circuit with the pruned Cuccaro adder

Stock price $A$ is encoded as an integer across multiple qubits, where each qubit represents a weighted component (e.g., 1x, 2x, 4x, 8x, ...). The proposed DS circuit is then employed to coherently aggregate these multi-qubit representations into a single qubit according to the corresponding weights. The resulting Boolean overflow flag serves as an indicator of the aggregated signal encoded in this single qubit. Functionally, the DS circuit serves as an agnostic, modular Oracle subroutine ($S_x$) within the standard QAE framework. During the expectation value estimation process, the Oracle conditionally activates the target qubit strictly based on the state of the Boolean overflow flag. This encapsulation allows QAE to efficiently extract the final weighted average with a quantum convergence rate of $O(1/m)$, fundamentally outperforming the classical direct measurement scaling of $O(1/\sqrt{m})$.

## IV. RESULT AND DISCUSSION

To provide a comprehensive evaluation of the proposed DS (Digital Spreading) framework against state-of-the-art benchmarks, this study analyzes its performance using an option pricing model based on the random walk model [1].

The experimental validation strictly utilized a discretized Black-Scholes-Merton (BSM) log-normal distribution to model the asset's random walk. The continuous probability density function for the spot price at maturity $S_T$ is defined as:

$$P(S_T) = \frac{1}{S_T \, \sigma \, \sqrt{2\pi T}} \exp\left(-\frac{(\ln S_T - \mu)^2}{2\sigma^2 \, T}\right)$$

where the drift term μ is given by $\mu = \ln(S_0) + \left(r - \frac{\sigma^2}{2}\right)T$. In our simulation, the financial parameters were configured with an initial price $S_0 = 2.0$, volatility $S_{T,i} = 0.10$, risk-free interest rate $r = 0.04$, and maturity $T = 300/365$ years.

To encode this distribution into a 3-qubit quantum register ($N = 8$ states), the continuous PMF was evaluated at eight uniformly spaced target prices $S_{T,i}$. The resulting probabilities were then strictly normalized to ensure unitarity:

$$\tilde{P}(S_{T,i}) = \frac{P(S_{T,i})}{\sum_{j=0}^{2^n - 1} P(S_{T,j})}$$

These normalized discrete probabilities, $\tilde{P}(S_{T,i})$, were mapped into the quantum amplitudes via exact state preparation, serving as the input state for the DS circuit.

To accommodate the constraints of realistic probability mass functions (PMFs), the probabilities generated by the random walk process are normalized, as illustrated in Fig. 9.

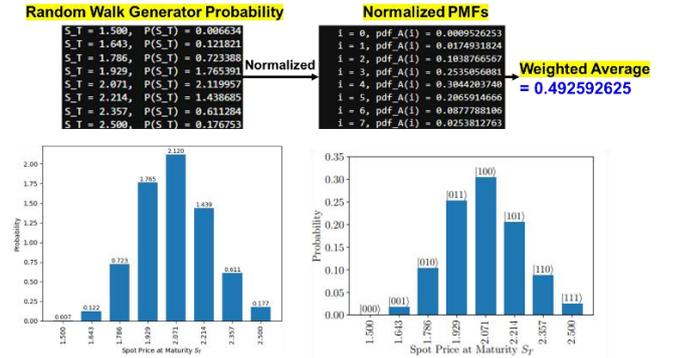

**Fig. 9**. The expected weighted average is 0.492592625 based on the option pricing model

The operational error—defined as the relative deviation of the final expected value from the theoretical ground truth—was analyzed using the normalized PMFs. As summarized in ***Table II***, the DS framework achieved an expected value of 0.492592988, exhibiting a relative error of merely 0.0001% compared to the 0.492592625 ground truth. This represents a statistically significant improvement over analog rotation-based methods (0.1679% error) and standard digital adders like QWA (0.0315% error at $10^6$ shots).

***Table II*** *Weighted Average and Relative Error*

| Method | Weighted Average | Error (%) |
|---|---|---|
| Expected Weighted Average | 0.492592625 | 0 |
| Digital Spreading | 0.492592988 | 0.0001 |
| QWA | 0.492748250 | 0.0315 |
| Taylor approximation | 0.491921138 | 0.1679 |

This performance represents a statistically significant improvement over incumbent state-of-the-art methodologies:

**1) Analog Rotation-Based Methods** (e.g., J.P. Morgan)**:** These approaches exhibited a substantially higher relative error of 0.1679%. The discrepancy is primarily attributed to inherent "sine-to-square" biases introduced by first-order Taylor series approximations.

**2) Standard Digital Adders** (e.g., QWA): Even under high-sampling conditions (one million measurement shots), the QWA method recorded a relative error of 0.0315%. These findings confirm that DS successfully eradicates the systematic errors plaguing analog methods while outperforming standard digital counterparts in convergence efficiency.

In summary, we propose a new framework for quantum expectation computation that operates without rotation gates or explicit arithmetic circuits. Previous rotation-based approaches rely on Taylor approximations, which, as discussed above, require subtracting 0.5 in the measured probability, leading to increased relative errors. Similarly, the conventional QWA requires lots of gates to perform explicit arithmetic operations. In contrast, the proposed DS framework offers two key advantages: enhanced error mitigation and reduced circuit complexity.

## V. Conclusion

This study proposes and successfully validates Digital Spreading (DS), a novel, fully integrated quantum arithmetic framework for computing weighted averages of probability distributions encoded within quantum amplitudes. By shifting the paradigm from analog-style rotation approximations to a discrete digital framework, DS achieves a relative error of merely 0.0001%. This fundamentally overcomes the inherent approximation errors, sine-to-square biases, and error magnification found in state-of-the-art rotation methods, such as those developed by JP Morgan and ITRI.

Furthermore, by directly mapping multi-qubit probability distributions onto a single target qubit, DS substantially reduces the computational overhead associated with the standard QWA. It requires fewer ancillary qubits and strictly confines circuit depth to a linear growth model, effectively mitigating decoherence risks on near-term hardware. This optimized, fully digital implementation streamlines the critical interface between classical data and quantum probability distributions. Ultimately, DS establishes a highly scalable and robust paradigm for high-precision Monte Carlo simulations. Beyond financial option pricing, this approach can accelerate expected-value computations across diverse disciplines—such as averaging molecular energies in chemistry, simulating particle systems in physics, or computing Bayesian posterior expectations in machine learning—delivering enhanced precision at a fraction of the traditional computational cost.